\documentclass[
 twocolumn,
]{ceurart}

\usepackage{algorithm}
\usepackage{algorithmic}

\usepackage{array}
\newcolumntype{L}[1]{>{\raggedright\let\newline\\\arraybackslash\hspace{0pt}}m{#1}}
\newcolumntype{C}[1]{>{\centering\let\newline\\\arraybackslash\hspace{0pt}}m{#1}}
\newcolumntype{R}[1]{>{\raggedleft\let\newline\\\arraybackslash\hspace{0pt}}m{#1}}

\usepackage{tabularx}

\usepackage{forest}       % For drawing the feature model tree
\usepackage{graphicx}     % For including images (if needed elsewhere)
\usepackage{float}        % For precise control over figure placement
\usepackage{amsmath}      % For math symbols (if needed in other sections)
\usepackage{hyperref}     % for refs

\usepackage{tikz}
\usetikzlibrary{trees}

\usepackage{listings}
\usepackage{newfloat}
\floatstyle{ruled}
\newfloat{listing}{tb}{lst}{}
\floatname{listing}{Listing}

\begin{document}

\copyrightyear{2025}
\conference{ConfWS'25: 27th International Workshop on Configuration, Oct 25--26, 2025, Bologna, Italy}

\title{Towards LLM-Enhanced Product Line Scoping }

% \tnotetext[1]{You can use this document as the template for preparing your
%   publication. We recommend using the latest version of the ceurart style.}

\author{Alexander Felfernig}[%
orcid=0000-0003-0108-3146,
email=alexander.felfernig@tugraz.at
]
% \cormark[1]
\fnmark[1]

\author{Damian Garber}[%
orcid=0009-0005-0993-0911,
email=damian.garber@tugraz.at
]
% \cormark[1]
\fnmark[1]

\author{Viet-Man Le}[%
orcid=0000-0001-5778-975X,
email=v.m.le@tugraz.at
]
% \cormark[1]
\fnmark[1]

\author{Sebastian Lubos}[%
orcid=0000-0002-5024-3786,
email=sebastian.lubos@tugraz.at
]
% \cormark[1]
\fnmark[1]

\author{Thi Ngoc Trang Tran}[%
orcid=0000-0002-3550-8352,
email=ttrang@ist.tugraz.at 
]
% \cormark[1]
\fnmark[1]

\address{Institute of Software Engineering and AI, Graz University of Technology, Graz, Austria}

% \cortext[1]{Corresponding author.}
\fntext[1]{These authors contributed equally.}

%%\renewcommand{\shortauthors}{Felfernig et al.}

%% paper.

\begin{abstract}
The idea of product line scoping is to identify the set of features and configurations that a  product line should include, i.e., offer for configuration purposes. In this context, a major scoping task is to find a balance between commercial relevance and technical feasibility. Traditional product line scoping approaches rely on formal feature models and require a manual analysis which can be quite time-consuming. In this paper, we sketch how Large Language Models (LLMs) can be applied to support product line scoping tasks with a natural language interaction based scoping process. Using a working example from the smarthome domain, we sketch how LLMs can be applied to  evaluate different feature model alternatives. We  discuss open research challenges regarding the integration of LLMs with product line scoping.
\end{abstract}


\maketitle

%%%%%%%%%%%%%%%%%%%%%%%%%%%%%%%%%%%%%%%%%%%%%%%%%%%%%%%%%%%%%%%%%%%%%%

\section{Introduction}

Configurable products and services such as smarthomes, cars, and software systems have a high variability in terms of which  components can be combined with each other \cite{featuremodelsbook2024,Felfernigetalknowledgebaseconfiguration2014}. To be able to handle variability in an efficient fashion,  product line (PL) approaches have been    widely adopted \cite{MetzgerPohlSoftwareProductLines2014}. The idea of product lines is to allow a systematic reuse of shared assets which enables the reduction of development costs, reduced time to market, and higher product quality. Product line scoping is at the heart of PL engineering \cite{deBaud1999,Johnetal2006,MarchezanetalSPLC2022,Ojedaetal2018,Schmid2002} -- it is the process of defining which features and constraints should be included in a product line, i.e., which features and corresponding constraints should be part of a feature model. High-quality scoping decisions are crucial  since they directly have an influence on the feasibility and commercial success of a product line.

Determining an optimal scope for a PL is a challenging task. This includes the evaluation of market trends, the balancing of potentially contradicting stakeholder requirements, and also ensuring the technical feasibility of the offered feature model configurations. A typical PL scoping process is based on workshops with experts. Related scoping decisions can be suboptimal due to a limited market and domain knowledge and -- as a consequence -- product lines have the risk of being under- or over-restricted. The underlying group  decision task makes product line scoping a task directly related to requirements prioritization \cite{LubosetalMeetingAnalysis2025,Sameretal2020} and group recommender systems \cite{Felfernigetal2024,LeetalMultistakeholderDecisionsSPLC2022,Masthoff2022}. 

Developments in the context of large language models (LLMs) \cite{Naveedetal2025LLMOverview} have created the potential to improve a variety of PL related tasks \cite{AcheretalProgrammingVariability2023,Greineretal2024LLMSPLCResearchVision,Hotzetal2024}. For example, in the context of software development, LLMs can be applied for re-engineering purposes allowing an LLM-based creation of PLs on the basis of artifacts such as UML diagrams, state charts, and Java programs \cite{AcherMartinez2023}. Furthermore, LLMs have shown to be applicable in the context of new feature identification from different textual sources such as  appstore evaluations \cite{Weietal2024}. Finally, LLMs have also shown to be applicable for model generation tasks, more precisely, the generation of feature models out of textual domain descriptions \cite{GalindoetalLLMs2023}. In the context of PL scoping, LLMs have the potential to support engineers in their tasks of analyzing trade-offs and identifying commercially promising variability concepts.

In this paper, we propose the idea of applying LLMs to pro-actively support different tasks in product line scoping. This includes the aspects of estimating model optimality (in terms of market potential, alignment with customer preferences, and cost efficiency) and technical feasibility of the offered features by taking into account the existing product development resources. 

We want to emphasize that these tasks are also relevant beyond  software product lines, for example, in the context of designing and configuring complex products such as cars and smarthome systems. As a basis for our discussions, we introduce a simplified working example from the domain of smarthome systems. With the introduced feature model, we sketch scenarios where LLM-supported product line scoping can provide help in estimating commercial relevance and technical feasibility.

The basic idea sketched in this paper is to exploit LLMs for pro-actively supporting group decision processes in product line scoping, i.e., we envision a scenario based on human-AI collaboration where LLMs provide additional insights (not covered by experts), indicate new alternatives, and explain the consequences of specific decisions \cite{Lubosetalconsequencesumap2025}. 

The major contributions of this paper are: first, we introduce the idea of exploiting LLMs for supporting product line scoping processes. Second, we sketch our ideas on the basis of a working example from the domain of smarthomes. Finally, we discuss  related open research issues.

The remainder of this paper is organized as follows. In Section 2, we present a feature model from the smarthome domain. In Section 3, we analyze different scenarios in which  LLMs can be employed to support product line scoping. Section 4 discusses open research issues. With Section 5, we conclude the paper.

\section{Working Example: SmartHome Feature Model}

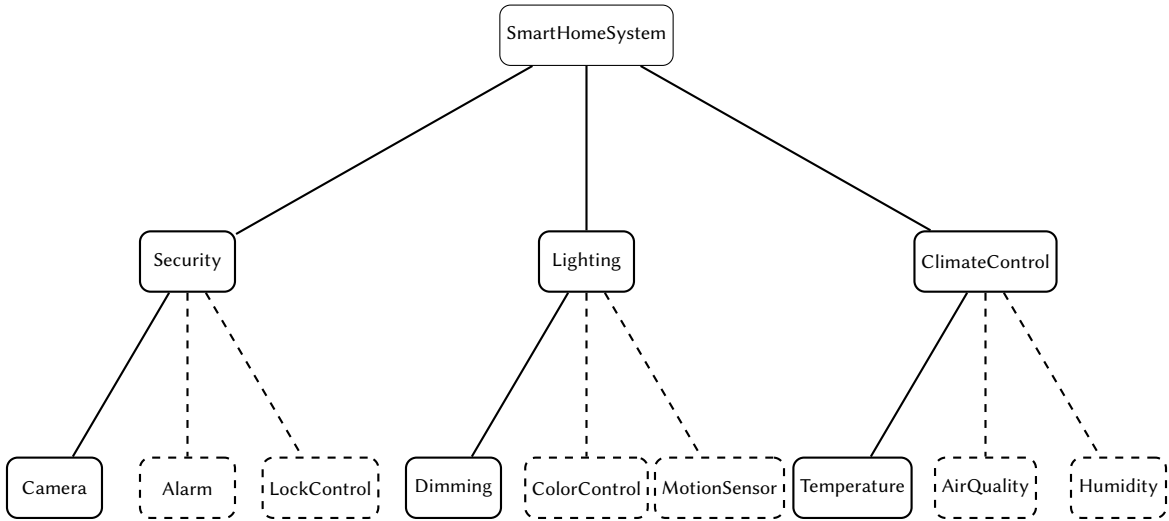
\begin{figure*}[ht]
\centering  \footnotesize
\begin{tikzpicture}[
  feature/.style={rectangle,draw,rounded corners,minimum width=1.25cm,minimum height=0.8cm,align=center},
  mandatory edge/.style={thick},
  optional edge/.style={dashed,thick},
  group edge/.style={thick,->},
  level 1/.style={level distance=3.0cm,sibling distance=5.25cm},
  level 2/.style={level distance=3.0cm,sibling distance=1.75cm},
  edge from parent/.style={draw,thick}
]

\node[feature] {SmartHomeSystem}
  child [mandatory edge] { node[feature] {Security}
    child [mandatory edge] { node[feature] {Camera} }
    child [optional edge] { node[feature] {Alarm} }
    child [optional edge] { node[feature] {LockControl} }
  }
  child [mandatory edge] { node[feature] {Lighting}
    child [mandatory edge] { node[feature] {Dimming} }
    child [optional edge] { node[feature] {ColorControl} }
    child [optional edge] { node[feature] {MotionSensor} }
  }
  child [mandatory edge] { node[feature] {ClimateControl}
    child [mandatory edge] { node[feature] {Temperature} }
    child [optional edge] { node[feature] {AirQuality} }
    child [optional edge] { node[feature] {Humidity} }
  };

\end{tikzpicture}
\caption{Feature model of a smarthome product line (dashed lines represent optional features, the other  mandatory ones.)} \vspace{0.25cm}
\label{fig:smart-home-feature-model}
\end{figure*}

In the following, we introduce a simplified feature model from the smarthome domain which will be used as a working example throughout the paper. Smarthome systems include a diverse set of features/functionalities including security, lighting, and climate control. Figure~\ref{fig:smart-home-feature-model} depicts a feature model of a  simplified smarthome product line.  The root feature \texttt{SmartHomeSystem} includes three basic subfeatures which are \texttt{Security}, \texttt{Lighting}, and \texttt{ClimateControl}. Each of those features has further subfeatures (either optional or mandatory ones).

In the feature model of Figure~\ref{fig:smart-home-feature-model}, the feature \texttt{SmartHomeSystem} is regarded as mandatory (due to the fact that we are not interested in empty configurations). The first-level child features \texttt{Security}, \texttt{Lighting}, and \texttt{ClimateControl} are represented as \emph{mandatory} which means that each smarthome configuration must include (in one way or another) each of those subfeatures (core features). Importantly, within the scope of a product line scoping process, this model can be regarded as flexible, i.e., features can be deleted or adapted and additional features  (and also constraints) can be included. Such adaptations can be triggered by new insights from market analyses as well as insights directly related to the technical feasibility of allowed configurations.

Features at the second level can either be mandatory of optional -- the features \texttt{Camera}, \texttt{Dimming}, and \texttt{Temperature} are regarded as mandatory, since they represent basic equipment to be included in every smarthome configuration. The remaining features of the model are regarded as optional, for example, the feature \texttt{ColorControl} can be offered to a customer but does not have to be included in every configuration. Note that further modeling concepts can be used for representing feature model properties. For a detailed discussion of feature model knowledge representations, we refer to Felfernig et al. \cite{featuremodelsbook2024,BeSeRu2010}.

Beyond hierarchical relationships such as mandatory and optional features, feature models often include cross-tree constraints that express dependencies between features. Such constraints further restrict the configuration space. Related example constraints in the smarthome domain could be \texttt{Alarm} requires \texttt{MotionSensor} (i.e., $Alarm \rightarrow MotionSensor$) and  \texttt{Alarm} excludes \texttt{ColorControl} (i.e., $\neg(Alarm \land ColorControl)$). On the basis of such a variability model, users can perform different analysis operations (representing individual queries on the feature model). Examples of such queries are: \emph{What are the minimum features required for a basic smarthome system with climate control}? or \emph{Which feature combinations are most relevant for urban apartment customers}? A related LLM-based assistant has the potential to provide explanations why specific features should be included in the feature model.

\section{Leveraging Large Language Models for Product Line Scoping}

Large language models (LLMs) have a deep contextual understanding and vast commonsense knowledge which makes them applicable in assisting complex decision-making. In the following, we analyze in which way LLMs can be used to analyse the optimality of a product line (in terms of market relevance and technical feasibility). Product line scoping includes the task of identifying which features or feature combinations are commercially relevant and technically feasible. In contrast to often manual scoping operations on the basis of feature models, LLMs can augment and partly automate scoping operations by supporting analysis operations, feasibility checks, and related commercial insights. 

In such scenarios, LLMs can be applied to answer scoping questions such as \emph{Does a smarthome system including \texttt{Security} with \texttt{Alarm} and \texttt{LockControl} but excluding \texttt{Lighting} make commercial sense?} In this example, an LLM can reason about potential consequences of omitting the \texttt{Lighting} feature. Furthermore, the related market acceptance could be estimated on the basis of knowledge about typical customer preferences, market trends in the smarthome domain, and technical background knowledge about the feasibility of such configurations. To some extent, LLMs can also take over reasoning tasks such as assuring that constraints integrated in the feature model do not induce an inconsistency.

Since LLMs are capable of processing natural language, they can be applied for developing conversational interfaces that support, for example, product line scoping processes. On the basis of such interfaces, users (members of the product line scoping team) can express complex queries without necessarily being able to understand the formal semantics of feature models. Furthermore, product line scoping is not necessarily based on feature models but can also be based on a textual definition of the variability properties of a product line (a blueprint-based representation). 

Examples of complex user queries are the following: a product manager might ask \emph{What is the most commercially attractive combination of security features for urban apartments}? or \emph{Suggest configurations that maximize energy efficiency while keeping costs low.}, or \emph{Create a feature model that supports the previously mentioned configurations.}

Such a query-based interaction in product line scoping is based on the following LLM-related capabilities. On the basis of available product domain knowledge, LLMs can identify when specific feature model parts are potentially triggering technical infeasibility. On the basis of information from product reviews, market reports, and other (potentially external) information sources in the training data, LLMs can estimate market potentials and the market relevance of specific features. 

LLMs are able to generate human-readable explanations of the provided feedback/explanations which can also be tailored to the users' background knowledge \cite{LubosetalLLMgeneratedExplanationsRecSys2024}. For example, technical argumentations can be provided to users with the corresponding technical background. This also helps to create transparency and decision confidence for users part of the product line scoping team. Furthermore, product line scoping can be supported in an interactive fashion, i.e., alternative feature model implementations can be explored and users receive immediate feedback on the implications of their current scoping decisions (model choices).

\section{Open Research Issues}

In the context of applying LLMs for supporting product line scoping processes, there exist a couple of open research issues which will be discussed in the following.

\vspace{0.2cm}

\emph{Reliability of LLM Feedback}. LLM feedback/assessment quality regarding product line optimality and feasibility needs to be assured. Since LLMs do not have formal reasoning capabilities, hallucinations and inconsistencies can occur (e.g., an LLM could generate feature models or parts thereof which are inconsistent, i.e., do not allow the generation of a configuration). In this context, hybrid approaches need to be developed which allow a combination of LLMs with formal consistency checking (e.g., on the basis of constraint solvers or SAT solvers) \cite{Hotzetal2024}. 

\vspace{0.2cm}

\emph{LLM Updates}. LLMs are based on domain-specific knowledge which experiences frequent updates. Since product line scoping has to depend on up-to-date market trend information and information about technological advances, and regulatory changes, efficient methods are needed that are able to continuously update the used LLMs. Furthermore, methods need to be developed that help to explain LLM feedback in terms of explaining the knowledge sources responsible for the given LLM feedback. This will help to support the aspects of transparency and trust which are crucial in the context of making high-involvement decisions for complex products and services. 

\vspace{0.2cm}

\emph{Scalability of Reasoning Services}. Since variability models can become quite large, the corresponding analysis and reasoning tasks require significant computational resources. Consequently, there is a need for reasoning services  within reasonable runtime performance.

\vspace{0.2cm}

\emph{Dialog Management}. Product line scoping is a complex (often group-based) decision task. This requires guidance in terms of proposing appropriate decision strategies to be used for completing a decision task and also in terms of informing the user in an understandable fashion about the next steps to be completed to achieve the overall goal of identifying an optimal variability model of a product line. In this context, natural language interaction can be quite intuitive for users. However,  communication has to be personalized, i.e., each user should receive system feedback and explanations in an understandable fashion.

\vspace{0.2cm}

 \emph{Sustainability Aspects}. Technical feasibility (T) and market relevance (M) are regarded as important decision criteria in the context of product line scoping. However, an important additional aspect to be taken into account are sustainability criteria (S) as defined by the United Nations Sustainability Development Goals (SDGs).\footnote{https://sdgs.un.org/goals} In this context, T, M, and C goals can be regarded as basic input of an optimization problem with the goal to identify optimal solutions.

\vspace{0.2cm}

\emph{Evaluation Metrics}. Evaluation metrics need to be developed that help to evaluate the outcomes of LLM-enhanced product line scoping processes. Related results need to be compared with the outcome of baseline processes without the support of LLM features. Example metrics include aspects such as commercial impact, technical feasibility, satisfaction of the customer community, and longterm positive sustainability effects.

\vspace{0.2cm}

\emph{Acceptance of Group Decision Support}. For different reasons, group decision support tools often suffer from limited user acceptance \cite{OjedaetalScoping2019}. On the one hand, such tools often require user feedback in terms of specifying explicit preferences which is not appreciated in complex scenarios such as product line scoping. On the other hand, there are issues related to aspects such as decision manipulation and limited preparedness to share his/her preferences. An important open issue in the context is find better ways of providing user support leading to more tool support acceptance as it is the case now.

\section{Conclusions}

In this paper, we have introduce the basic idea of exploiting large language models (LLMs) to the support decision processes in product line scoping for complex products and services. On the basis of a working example from the domain of smarthomes, we have sketched how variability modeling can be combined with LLMs with the goal to increase the quality of product line scoping. This way, stakeholders can be supported and guided in complex decision tasks in a more efficient fashion. 

However, there are a couple of open research issues including for example, the aspects of LLM feedback reliability and explainability of the LLM output. Our next steps will be a more detailed analysis of the commercial needs of LLM-supported product line scoping. The corresponding results will be the major features of our envisioned tool for supporting LLM-enhanced product line scoping.

%\begin{acknowledgments}
%The work presented in this paper has been conducted within the scope of the \textsc{OpenSpace} project funded by the Austrian research promotion agency (FO999891127). 
%\end{acknowledgments}

% \appendix

\bibliography{bibliography}

\end{document}